\documentclass[a4paper, 10pt, conference]{ieeeconf}      

\IEEEoverridecommandlockouts                              
\usepackage{cite}
\usepackage{graphicx}
\usepackage{amsmath}
\usepackage{multirow}
\usepackage{flushend}
\usepackage{xcolor}

\usepackage{draftwatermark}
\SetWatermarkText{Accepted by SusTech 2021 (an IEEE Conference)}
\SetWatermarkColor[gray]{0.1}
\SetWatermarkFontSize{0.75cm}
\SetWatermarkAngle{90}
\SetWatermarkHorCenter{20.5cm}
\usepackage[ruled,vlined]{algorithm2e}

\def\BibTeX{{\rm B\kern-.05em{\sc i\kern-.025em b}\kern-.08em
    T\kern-.1667em\lower.7ex\hbox{E}\kern-.125emX}}

\title{\Huge A GA-based Approach to Eco-driving of Electric Vehicles Considering Regenerative Braking}
\author{Mukesh Gautam, \emph{Student Member, IEEE}, Narayan Bhusal, \emph{Student Member, IEEE},\\ Mohammed Benidris, \emph{Member, IEEE}, and Poria Fajri, \emph{Member, IEEE}, \\
Department of Electrical and Biomedical Engineering, 
University of Nevada, Reno, NV 89557, USA\\
Emails: mukesh.gautam@nevada.unr.edu, bhusalnarayan62@nevada.unr.edu, \\ mbenidris@unr.edu, and pfajri@unr.edu\vspace{-3ex}}

\begin{document}
\maketitle
\thispagestyle{empty}
\pagestyle{empty}

\begin{abstract}
As the deployment of low carbon transportation technologies, specifically electric vehicles (EVs), is increasing, the concept of their eco-driving is gaining significant attention. Contrary to the eco-driving techniques used in conventional internal combustion engine vehicles that do not have the capability of regenerative braking, this paper proposes a genetic algorithm (GA)-based eco-driving technique for EVs considering regenerative braking. In the proposed approach, the optimal or near-optimal combination of variables in the driving cycle of EVs is searched using GA. The proposed approach starts by generating an initial population of chromosomes, where all variables under consideration are encoded in each chromosome. This population of chromosomes is passed through crossover, mutation, and elitist-based selection over a certain number of generations, which results in a driving cycle with the least energy consumption. The proposed method is verified using two case studies. The results of the case studies show the capability of the proposed method in computing the minimum energy driving cycle.\vspace{-1ex}   
\end{abstract}
\begin{keywords}
Eco-driving, electric vehicle, genetic algorithm, and regenerative braking\vspace{-0.8ex}. 
\end{keywords}

\section{Introduction}
Deployment of low carbon transportation technologies, specifically electric vehicles (EVs), is increasing worldwide in recent years. Several factors have contributed in the continuous sharp increase of production and use of EVs including environmental concerns, governmental incentives (e.g., tax credit), limited supply of fossil fuels, efficiency improvement of EVs and battery technologies, and the decrease in the production cost of EVs and batteries \cite{NB2018thesis}. However, this growth is still not in the scale of conventional internal combustion vehicles. According to the InsideEVs \cite{INSIDEEVs}, only $1$ in $40$ new vehicles sold is an EV. Numerous customers are hesitant to use EV mainly due to limited battery capacity and long charging time. This limits the travel time and distance of vehicles, sometimes referred to as ``range anxiety.'' 
The regenerative braking can serve as an option to recover the energy during braking, which extends the driving range of EV. However, the dependency of regenerative braking performance of an EV on various factors influenced by the driver behavior and driving conditions makes the problem computationally expensive \cite{heydari2020optimal}.
Therefore, this calls for a computationally efficient and accurate approach that can reduce the energy consumption by EVs resulting in an increase in driving range. 

As the energy conversion in an EV passes through several steps while being on the road, different energy-saving techniques can be employed in each step. The three main energy conversion steps (grid-to-tank, tank-to-wheels, and wheels-to-distance) for a road vehicle have been stated in \cite{8378052}. Out of these three energy steps, the energy conservation in wheels-to-distance is the main focus of this paper. The energy efficiency at the wheel-to-distance stage can be improved by controlling the driving profile, commonly referred to as ``eco-driving.'' Eco-driving refers to driving in such a way that the total energy consumption during a driving cycle is minimized. The optimal running profile of a vehicle is obtained as a result of eco-driving techniques. The eco-driving techniques are highly essential because they not only increase the battery life cycle but also the vehicle range.

Numerous approaches have been proposed in the literature for solving the eco-driving problem in EVs. A global optimal solution to eco-driving problem using sequential quadratic programming algorithm has been presented in \cite{8378052}. The solution of eco-driving problem as an optimal control problem has been suggested in \cite{7265166}. The speed advisory system for the minimization of energy consumption of vehicle over a planned route as an optimal control problem has been presented in \cite{8286942}. In \cite{dongreoptimization}, the energy consumption in an electric traction system has been optimized using interior point algorithm. In \cite{PAREDES2019556}, a shrinking horizon approach has been proposed for eco-driving of electric city buses. Authors of \cite{8703854} have proposed receding horizon and optimal control strategy to compute eco-driving cycle for electric, conventional, and hybrid vehicles. The problem of eco-driving has been formulated as consecutive optimization problems aiming at minimizing the vehicle energy consumption under traffic and speed constraints. In \cite{KIM2020114254}, Pontryangin's minimum principle based co-optimization strategy of speed trajectory and power management has been proposed for fuel-cell/battery electric vehicle. In \cite{9045962}, an optimal speed profile has been computed for eco-driving using constant pedal position technique. In \cite{7115224}, an eco-driving at signalized intersections for EV with the support of vehicle infrastructure system technologies has been proposed. A unified approach for EV range maximization via eco-routing, eco-driving, and energy consumption prediction using artificial neural network has been presented in \cite{8481449}.

Although several methods of eco-driving of EVs have been proposed in the exiting literature, the consideration of regenerative braking, as an effective way to recover the energy that would otherwise be wasted, in eco-driving has received little attention. As the purpose of the eco-driving is to minimize the energy consumption subject to vehicle constraints, regenerative braking recovers the braking energy that results in the energy saving.

This paper proposes a GA based approach for eco-driving of EVs. The proposed GA based method starts by reading vehicle data and generating an initial population of chromosomes, where all variables under consideration are encoded in each chromosome. This population of chromosomes is passed through crossover, mutation, and elitist-based selection. 
During elitist-based selection process, the total energy consumption for a driving cycle represented by each chromosome is computed. The total energy consumption is computed by subtracting the regenerative energy from the energy consumed during acceleration and constant cruising. After evaluating this over a certain number of generations, a driving cycle with the least amount of energy consumption is obtained.

The remainder of the paper is organized as follows. The modeling of an electric vehicle including its dynamics is described in section \ref{model}. Section \ref{prob} describes the formulation of the optimization method for an eco-driving problem. Section \ref{methods} presents the proposed methodology for the implementation of genetic algorithm based eco-driving technique. Section \ref{results} validates the proposed approach with case studies and discussions. Section \ref{conclusion} provides concluding remarks.  

\section{System Modeling}\label{model}
In order to compute the energy consumption of an EV, it is necessary to model the vehicle dynamics and the whole power-train from battery to the wheels. The longitudinal dynamics of vehicle can be expressed as follows.
\begin{equation}
    M\frac{dV}{dt}=F_t-F_r \label{motion}
\end{equation}
where $M$ is vehicle mass; $V$ is vehicle speed; $\frac{dV}{dt}$ denotes vehicle acceleration; $F_t$ is the total tractive force; and $F_r$ denotes the total resistive force experienced by the vehicle. The total resistive force is assumed to consist of rolling resistance and aerodynamic drag only.

Considering rolling resistance and aerodynamic drag, the total tractive power needed to accelerate a vehicle from zero to speed $V_c$ in $t_a$ seconds can be expressed as follows \cite{ehsani2018modern}.
\begin{equation}
    P_a=\frac{\delta M}{2t_a}(V_c^2+V_b^2)+\frac{2}{3}Mgf_rV_c +\frac{1}{5}\rho_a C_d A_f V_c^3 \label{Pa}
\end{equation}
where $\delta$ is the vehicle mass factor; $g$ is the acceleration due to gravity; $f_r$ is the rolling resistance coefficient; $\rho_a$ is the air density; $C_d$ is the aerodynamic drag coefficient; and $A_f$ is the vehicle frontal area.
\begin{figure}
    \hspace{-2ex}
    \includegraphics[scale=0.25]{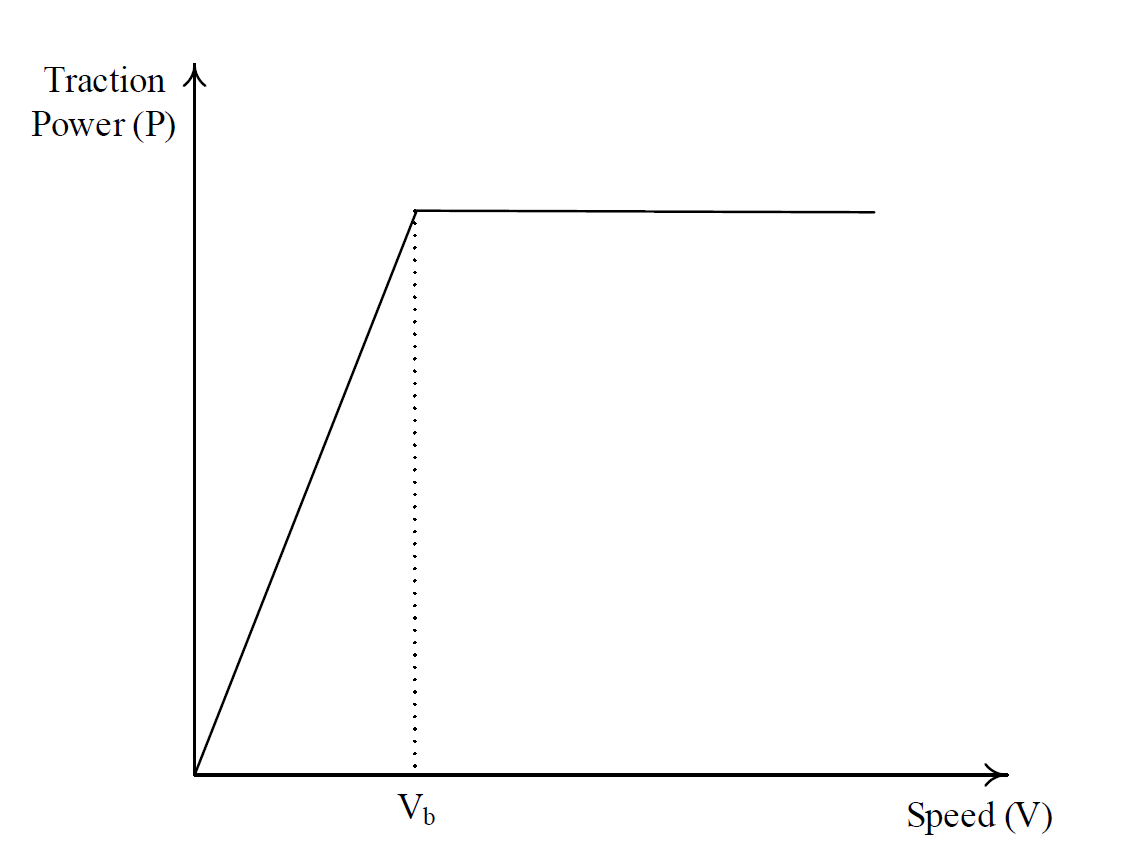}
    \vspace{-2ex}
     \caption{Power-speed characteristic of a typical traction drive}
    \label{Powerspeedchar}
    \vspace{-2ex}
\end{figure}

In \eqref{Pa}, $V_b$ is the base speed of traction drive, indicated in Fig. \ref{Powerspeedchar}, which is related to the speed ratio $x$ and the maximum cruising speed $V_{max}$ as follows.
\begin{equation}
    x=\frac{V_{max}}{V_b}
\end{equation}

Similarly, the traction power during constant cruising is computed as follows.
\begin{equation}
    P_c=Mgf_rV_c +\frac{1}{2}\rho_a C_d A_f V_c^3 \label{Pc}
\end{equation}

During braking (or retardation) from speed $V_c$ to zero in $t_b$ seconds, the required braking power can be computed as follows.
\begin{equation}
    P_b=\frac{\delta M}{2t_b}(V_c^2+V_b^2)+\frac{2}{3}Mgf_rV_c +\frac{1}{5}\rho_a C_d A_f V_c^3 \label{Pb}
\end{equation}

\section{Problem Setting}\label{prob}
The main purpose of eco-driving technique is to minimize the total energy consumption for a driving cycle subject to vehicle dynamics and driving constraints. The optimization problem for the proposed eco-driving technique can be formulated as follows.
\begin{equation}
    \text{min} \{E=\int_0^T V(t) F_w(t)dt \} \label{obj}
\end{equation}
subject to
\begin{equation}
    \int_0^T V(t) dt =S \label{dist}
\end{equation}
\begin{equation}
    V(t) \leq V_{max} \label{maxv}
\end{equation}
\begin{equation}
    a \leq a_{max} \label{maxa}
\end{equation}
\begin{equation}
    T \leq T_{max} \label{maxt}
\end{equation}

where \eqref{obj} calculates total energy consumption $(E)$; \eqref{dist} is travel distance constraint with $S$ as total distance to be travelled; \eqref{maxv} represents the maximum speed limit constraint; \eqref{maxa} denotes the maximum acceleration limit constraint;  \eqref{maxt} is the maximum time limit constraint; $T$ is total travel time; $V(t)$ is the instantaneous speed at time $t$; $F_w(t)$ is the total traction force transmitted on to the wheels; $V_{max}$ is the maximum cruising speed limit; $a_{max}$ is the maximum acceleration limit; and $T_{max}$ is the maximum time limit.

\begin{figure}
    \centering
    \hspace{-3.2ex}
    \includegraphics[scale=0.25]{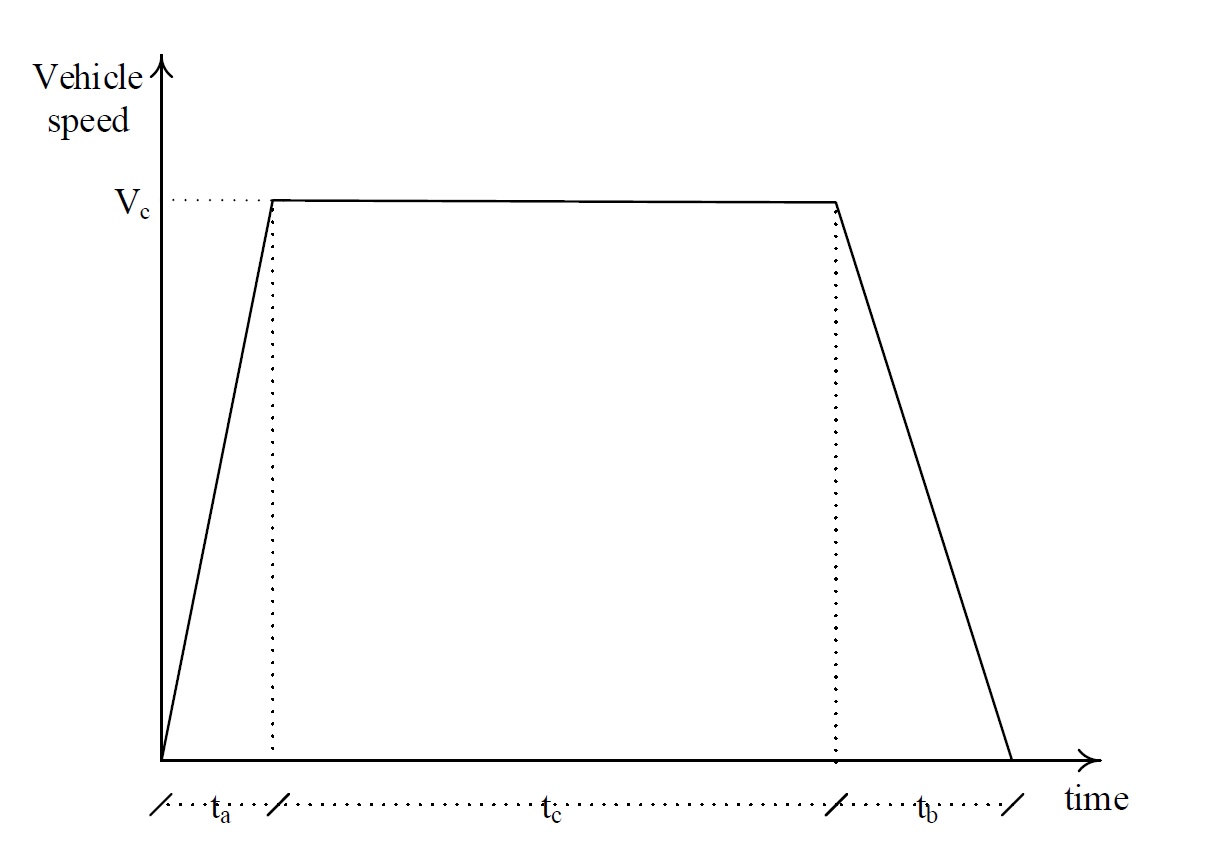}
    \vspace{-1ex}
    \caption{Speed Profile for a simple driving cycle (case $I$)}
    \label{profile_case1}
    \vspace{-1ex}
\end{figure}

For a simple driving cycle (case $I$) as shown in Fig.~\ref{profile_case1}, the optimization problem can be written as follows.

\begin{equation}
    \text{min} \{E= P_at_a+P_ct_c-P_bt_b\} \label{obj1}
\end{equation}
subject to
\begin{equation}
    \frac{1}{2}t_a V_c +t_c V_c + \frac{1}{2} t_b V_c =S \label{dist1}
\end{equation}
\begin{equation}
    V(t) \leq V_{max} \label{maxv1}
\end{equation}
\begin{equation}
    a \leq a_{max} \label{maxa1}
\end{equation}
\begin{equation}
    t_a+t_b+t_c \leq T_{max} \label{maxt1}
\end{equation}
where $t_a$ is accelerating time; $t_c$ is constant cruising time; and $t_b$ is braking time.

\begin{figure}
    \centering
    \hspace{-3.2ex}
    \includegraphics[scale=0.25]{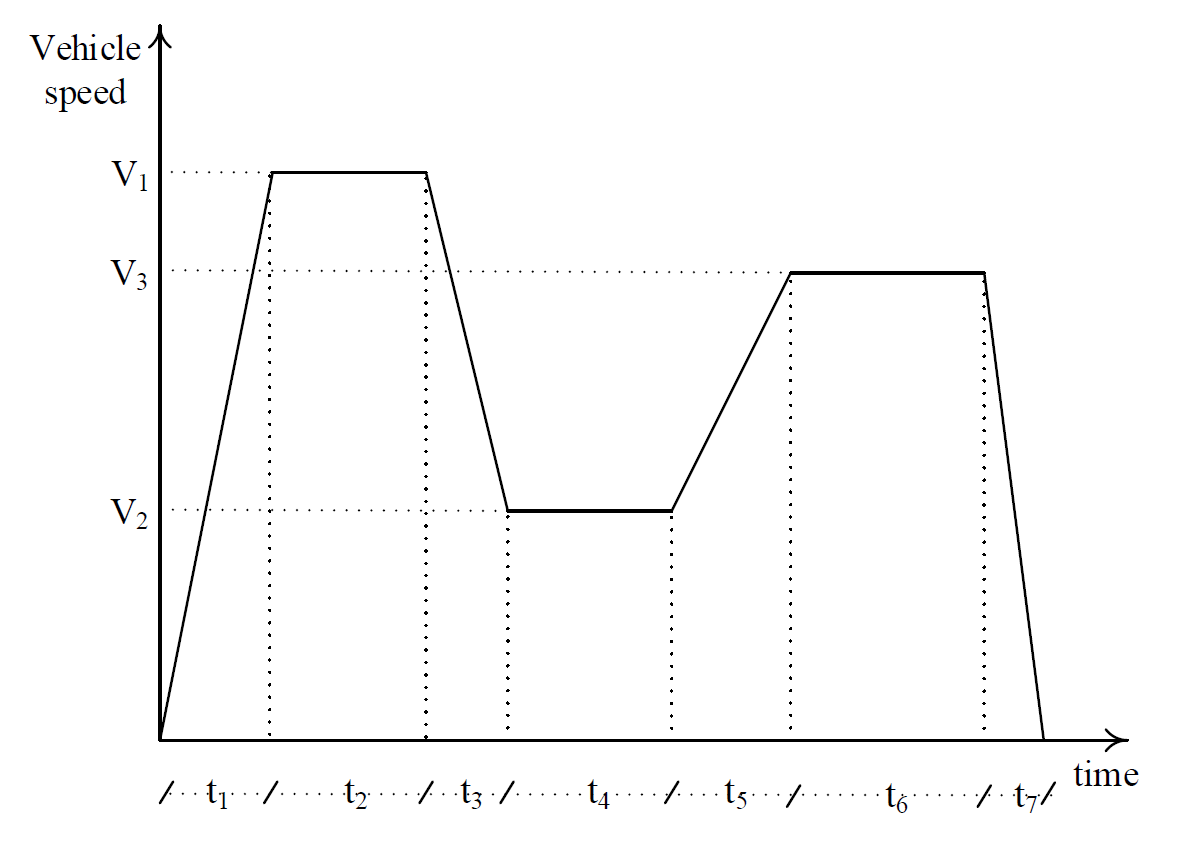}
    \vspace{-1ex}
    \caption{Speed Profile for case $II$}
    \label{profile_case2}
    \vspace{-2ex}
\end{figure}

For case $II$ (road with speed limits) having driving cycle as shown in Fig.~\ref{profile_case2}, the optimization problem can be formulated as follows.
\begin{equation}
    \text{min} \{E= P_1t_1+P_2t_2-P_3t_3+P_4t_4+P_5t_5+P_6t_6-P_7t_7 \}\label{obj2}
\end{equation}
subject to
\begin{equation}
   \int_0^T V(t) dt =S \label{dist2}
\end{equation}
\begin{equation}
    V(t) \leq V_{max} \label{maxv2}
\end{equation}
\begin{equation}
    a \leq a_{max} \label{maxa3}
\end{equation}
\begin{equation}
    t_1+t_2+t_3+t_4+t_5+t_6+t_7 \leq T_{max} \label{maxt1}
\end{equation}
where $P_1$, $P_2$, $P_3$, $P_4$, $P_5$, $P_6$, and $P_7$ are tractive powers for time intervals of $t_1$, $t_2$, $t_3$, $t_4$, $t_5$, $t_6$, and $t_7$ respectively.

The minimization type objective function in \eqref{obj} or \eqref{obj1} or \eqref{obj2} is converted into maximization fitness function $F$ as follows.
\begin{equation} \label{fitness}
    F = \frac{1}{1 + E}
\end{equation}

While solving the eco-driving problem, the optimization algorithm needs to search for various combinations of independent variables (acceleration, constant cruising speed, and retardation) to determine the optimal combination resulting in the minimization of the total energy consumption of the vehicle. If this problem is searched exhaustively for all possible combinations of independent variables, $2^n$ number of combinations are generated, where $n$ is the number of bits of each chromosome. For example, the search space is equal to $16,384$ for the simple drive cycle under consideration. As the complexity in the driving cycle increases, the search space becomes even larger. GA based search approach for optimal or near optimal solution has been successfully applied in various fields \cite{9067540, 265940,7764202,METAWA201775, HIASSAT201793, IAS2020mukesh}, therefore, it is adopted in this paper to solve the eco-driving problem to minimize the total energy consumption of EVs.

\section{Solution Methodology}\label{methods}
This section describes the solution method for the proposed GA-based approach. In the proposed work, the energy consumed by an EV over a driving cycle is minimized resulting in the generation of an optimal or near-optimal driving profile by taking regenerative braking into account. Contrary to the conventional vehicles with friction braking, the regenerative braking in EVs serves as an effective way to recover the energy that would otherwise be wasted in the form of heat. As already mentioned in the previous section that GA has been successfully applied in various fields to obtain optimal or near-optimal, this paper has adopted GA to solve the eco-driving problem taking regenerative braking into account.  

The effectiveness of genetic algorithm based optimization techniques highly depend on how the solutions are represented. The solutions of a GA should be represented in such a way that it is able to search for all possible solutions. 

\begin{figure}[h!]
    \centering
    \includegraphics[scale=0.2]{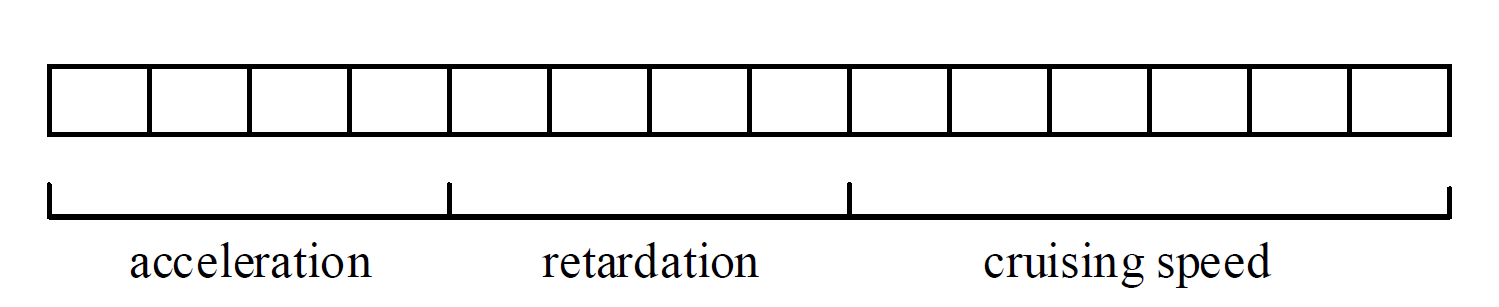}
    \vspace{-1ex}
    \caption{Solution Representation\vspace{-3ex}}
    \label{bit}
    \vspace{-1ex}
\end{figure}

In this work, all the independent variables under consideration (i.e.  rate of acceleration, cruising speed, and rate of retardation) are encoded in binary strings for solution representation. For example, for the case $I$, each individual (or chromosome) is encoded as a $14$-bit binary number with $4$ bits representing acceleration, $4$ bits representing retardation, and the remaining $6$ bits represent maximum speed, as shown in Fig.~\ref{bit}. Similarly, for case $II$, each chromosome is encoded as a string of $32$ binary digits. 

The three main GA operators utilized in this study are as follows.
\begin{itemize}
    \item Crossover: Single point crossover is used in this work. In a single-point crossover, a cross-over point is randomly selected and bits to the right of this point are swapped between parent chromosomes.
    \item Mutation: Single point bit-flip mutation is used in this work. In single-point bit-flip mutation, a mutation point is randomly selected and the bit at this point is flipped.
    \item Elitist Selection: Elites are individuals with the best fitness in current generation. In the selection methods other than elitist selection, there is a chance of elites being eliminated through cross-over and mutation operations. Therefore, in this work, the elitist selection is applied for preserving the best fit individuals, which are automatically passed to the next generation. 
\end{itemize} 

\begin{figure}
    \centering
    \hspace{-3.2ex}
    \includegraphics[scale=0.5]{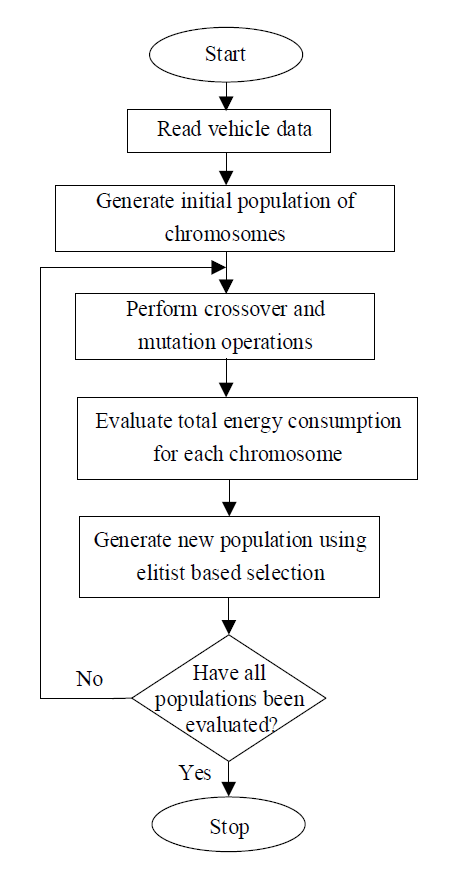}
    \vspace{-1ex}
    \caption{The flowchart of the proposed methodology.}
    \label{flowchart}
    \vspace{-1ex}
\end{figure}

During the elitist-based selection process, the total energy consumption for each driving cycle represented by an individual chromosome is computed. During retardation or braking phases of each driving cycle, a certain portion of the kinetic energy of the vehicle is recovered and is utilized in charging the battery.
The recovery or regeneration of a certain portion of the energy during braking results in energy-saving.
The flow-chart as shown in Fig.~\ref{flowchart} describes the proposed solution approach.

\section{Case Studies and Discussion}\label{results}
In this section, the proposed method is validated through two case studies. The case $I$ consists of a $5$-mile road with a simple driving cycle (similar to Fig.~\ref{profile_case1}, consisting of an acceleration, a constant cruising speed, and retardation). In case $II$, a $5$-mile road with a speed limitation in middle $1$-mile stretch is considered resulting in a driving cycle similar to Fig.~\ref{profile_case2}. The vehicle parameters and their values used for the case studies are shown in Table~\ref{tab:data}. For comparison, both the cases are also run using stochastic hill climber (SHC). The algorithm for SHC is as shown in Algorithm~\ref{alg}. The details of the case studies are as follows.
\begin{table}[h!]
\caption{Vehicle data used for case studies}
\centering
\label{tab:data}
\begin{tabular}{|c|c|}
\hline
\textbf{Parameters} & \textbf{Values} \\ \hline
Vehicle mass $(M)$                     & $2000$ kg                           \\ \hline
Air density $(\rho_a)$                      & $1.22$ kg/m$^3$                         \\ \hline
Aerodynamic drag coefficient $(C_d)$                     & $0.3$                          \\ \hline
Frontal Area $(A_f)$                       & $1.6$ m$^2$                         \\ \hline
Rolling resistance coefficient $(f_r)$                     & $0.01$                        \\ \hline
Wheel radius $(r_d)$                      & $0.28$ m                         \\ \hline
Wind speed $(V_w)$                    & $0$                          \\ \hline
Vehicle mass factor $(\delta)$                     & $1.04$                       \\ \hline
Gear ratio of final drive/differential $(i_0)$                       & $4.18$                         \\ \hline
Gear ratio of transmission $(i_g)$                   & $1.3$                            \\ \hline
Speed ratio, maximum speed to base speed $(x)$                   & $4$                            \\ \hline
Traction motor efficiency $(\eta_m)$                   & $85\%$                            \\ \hline
Inverter efficiency $(\eta_i)$                   & $95\%$                            \\ \hline
Gear-box and final drive efficiency $(\eta_t)$                   & $90\%$                            \\ \hline
Total travel distance $(S)$                   & $5$ miles                            \\ \hline
Maximum time limit $(T_{max})$                   & $420$ sec                      \\ \hline
Maximum acceleration $(a_{max})$                   & $8$ mph/s                      \\  \hline
Maximum cruising speed $(V_{max})$                   & $75$ mph                      \\ \hline
Overall regenerative efficiency                   & $50\%$                      \\ \hline
\end{tabular}
\vspace{-2ex}
\end{table}
\begin{algorithm}\label{alg}
\SetAlgoLined
\KwResult{Final solution $fSol$ }
 initialize a randomly generated solution $iSol$\;
 \For{$i = 1$ to $\text{max}_{\text{itr}}$}{
  generate new solution $nSol$\;
  \If{\text{fitness}($nSol$) $\geq$ fitness($iSol$)}{
   $iSol$:=$nSol$\;
   }{
  }
 }
 $fSol:=iSol$\;
 \caption{Stochastic Hill Climber}
\end{algorithm}\vspace{-2.5ex}

\subsection{Case $I$: Road with simple driving cycle}
In this case, a $5$-mile road is considered that does not have any speed limits and the speed may be limited only by the vehicle constraints. A simple driving cycle consisting of an acceleration ($\alpha$), a constant cruising speed ($V$), and retardation ($\beta$) is assumed in this case. In order to check the effectiveness of the proposed method, the results are compared with that of SHC.

\begin{figure}
\vspace{-2ex}
    \centering
    \hspace{-3.2ex}
    \includegraphics[scale=0.5]{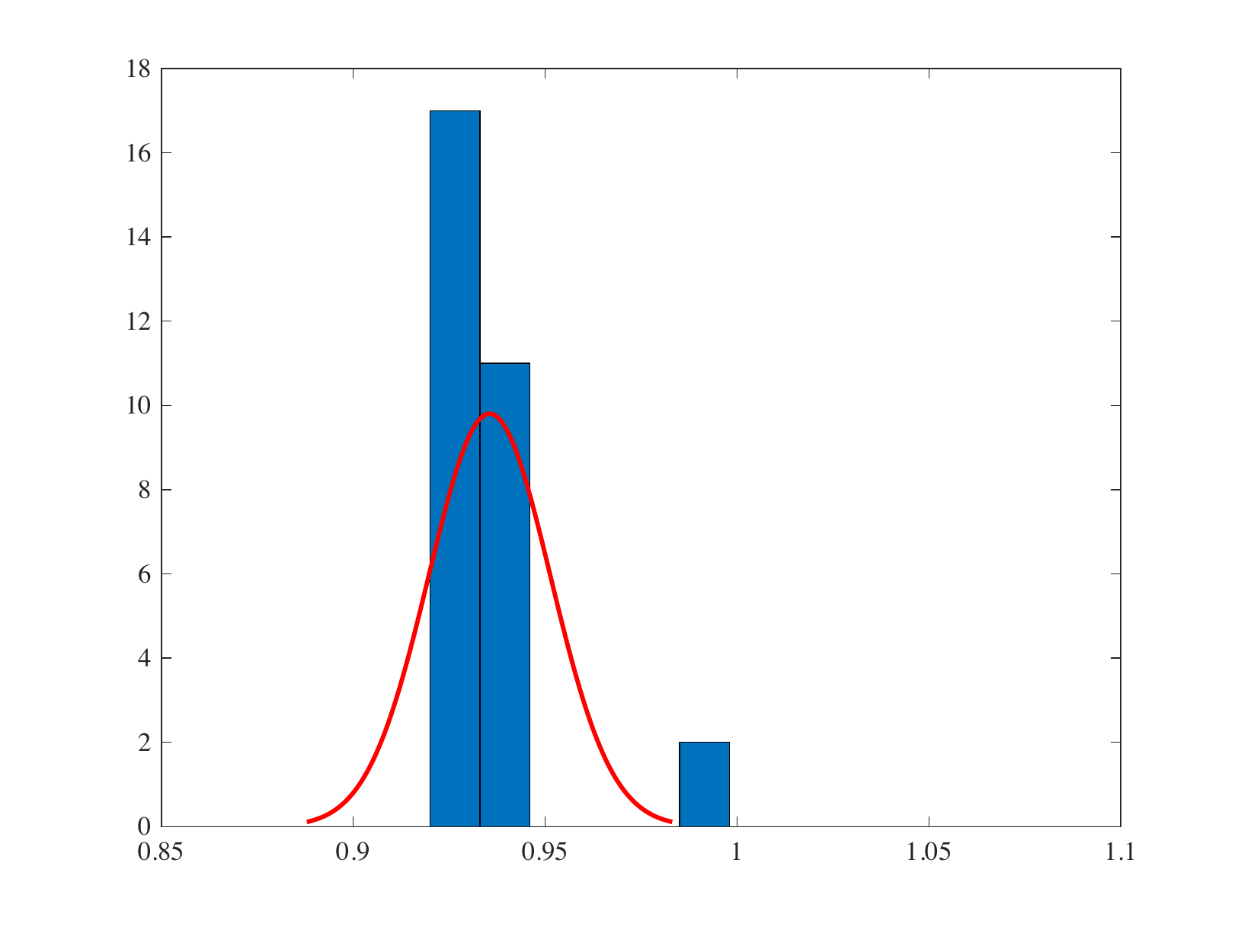}
    \vspace{-3ex}
    \caption{Histogram showing distribution of $E_{min}$ for $30$ runs of Case $I$ using proposed method}
    \label{hist_case1_GA}
\end{figure}

\begin{figure}
    \hspace{-3.2ex}
    \includegraphics[scale=0.5]{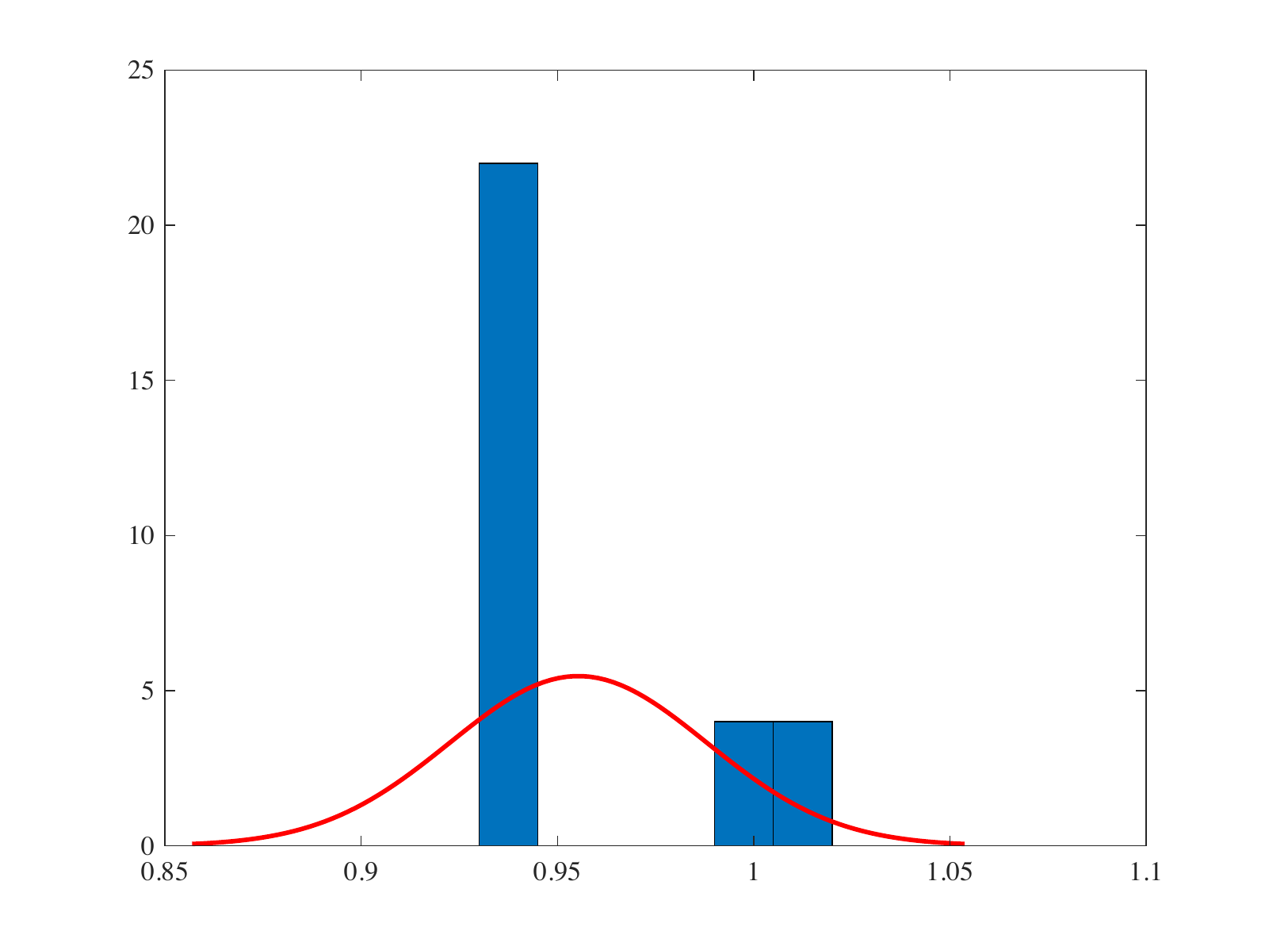}
    \vspace{-3ex}
    \caption{Histogram showing distribution of $E_{min}$ for $30$ runs of Case $I$ using stochastic hill climber}
    \label{hist_case1_HC}
    \vspace{-2ex}
\end{figure}

The optimized value of parameters, viz. acceleration ($\alpha$) in miles per hour per second (mph/s), constant cruising speed ($V_c$) in mile per hour (mph), and retardation ($\beta$) in mph/s for the proposed method and SHC are shown in Table~\ref{tab:Case1}. Table~\ref{tab:Case1} also shows the overall minimum energy consumption ($E_{min}$) and the average ($E_{avg}$) and standard deviation ($\sigma$) of minimum energy consumption over $30$ runs for both the methods. The results in the table show that SHC got stuck over local optimum with $E_{min}$ equal to $0.9361$ kWh whereas the proposed method could search for the better solution with $E_{min}$ equal to $0.9285$ kWh. The table also shows that the results obtained using SHC have higher variance than the proposed method. The histograms showing the distribution of minimum energy consumption ($E_{min}$) over $30$ runs for the proposed method and SHC are shown in Fig.~\ref{hist_case1_GA} and Fig.~\ref{hist_case1_HC} respectively.
\begin{table*}[h!]
\caption{Results for Case $I$ over 30 runs \vspace{-1.5ex}}
    \centering
        \label{tab:Case1}
    \begin{tabular}{|c|c|c|c|c|c|c|}
\hline
\multirow{2}{*}{Methods}  & \multicolumn{3}{c|}{Optimized   Parameters}                              & \multirow{2}{*}{$E_{min}$   (kWh)} & \multirow{2}{*}{$E_{avg}$ (kWh)} & \multirow{2}{*}{$\sigma$ (kWh)} \\ \cline{2-4}
                          & $\alpha$ (mph/s) & $\beta$ (mph/s) & $V$   (mph) &                                     &                                    &                                                \\ \hline
Stochastic   Hill Climber & 8                             & 0.5                          & 50.40     & 0.9361                              & 0.9553                             & 0.0328                                         \\
Proposed   Method         & 8                             & 0.5                          & 49.60     & 0.9285                              & 0.9355                             & 0.0159                                         \\ \hline
\end{tabular}\vspace{-1ex}
\end{table*}

\begin{figure}
    \centering
    \hspace{-3.2ex}
    \includegraphics[scale=0.5]{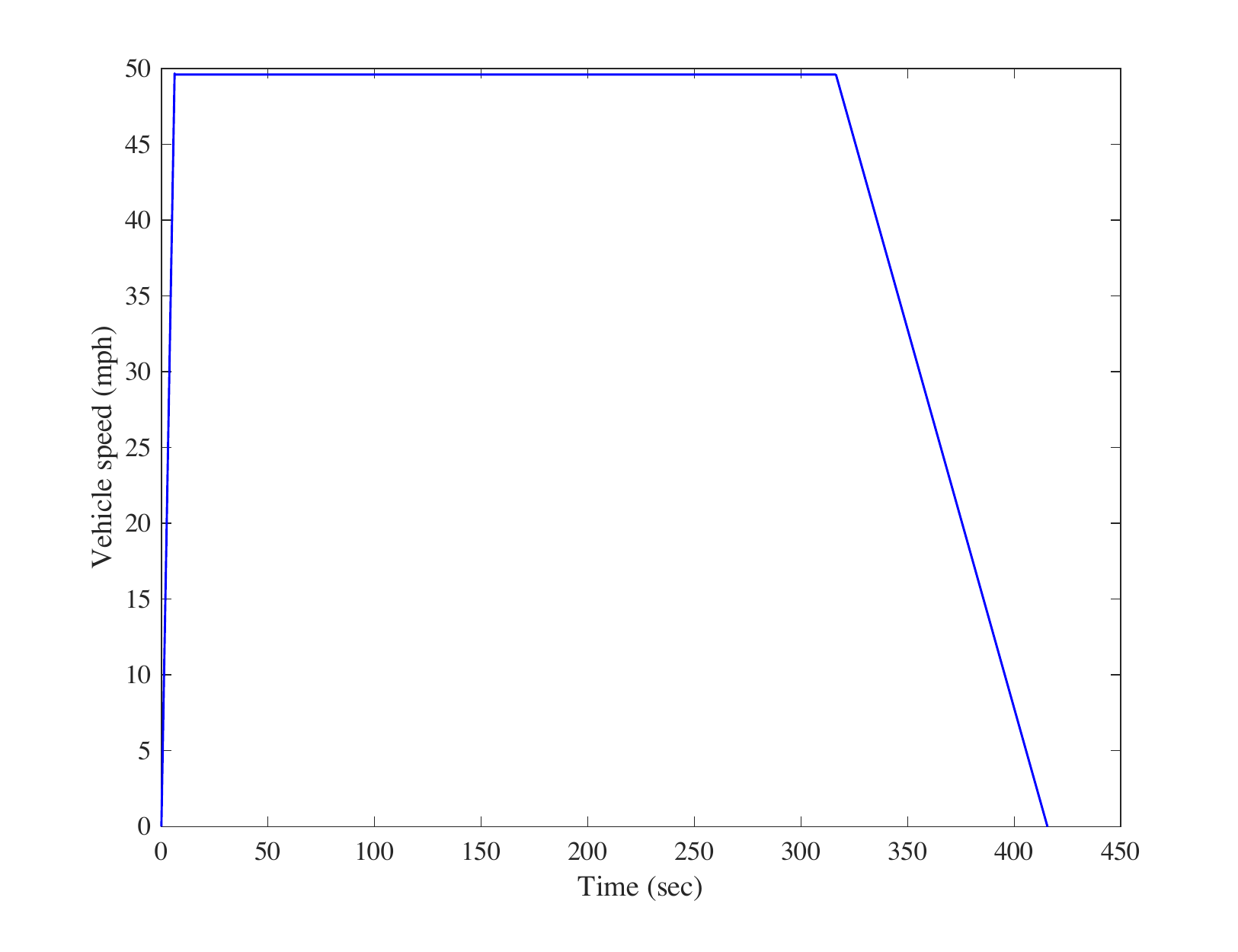}
    \vspace{-3ex}
    \caption{Optimal Speed Profile for Case $I$}
    \label{speedprofile_case1}
    \vspace{-2ex}
\end{figure}

The optimal speed profile obtained for this case using the proposed method is shown in Fig.~\ref{speedprofile_case1}. The figure clearly shows a high acceleration of $8$ mph/s, constant cruising speed $49.60$ mph, and a low retardation of $0.5$ mph/s.

\subsection{Case $II$: Road with speed limits}
In this case, a $5$-mile road is considered that consists of $1$-mile stretch in the middle with speed limits of $25$ mph. Because of the presence of $1$-mile stretch in the middle with speed limits, the driving cycle will consist of two accelerations $\alpha_1$ and $\alpha_2$, three constant cruising speeds $V_1$, $V_2$, and $V_3$, and two retardations $\beta_1$ and $\beta_2$. In order to check the effectiveness of the proposed method, the results of this case are also compared with that of SHC. 

\begin{figure}[h!]
    \centering
    \hspace{-3.2ex}
    \includegraphics[scale=0.5]{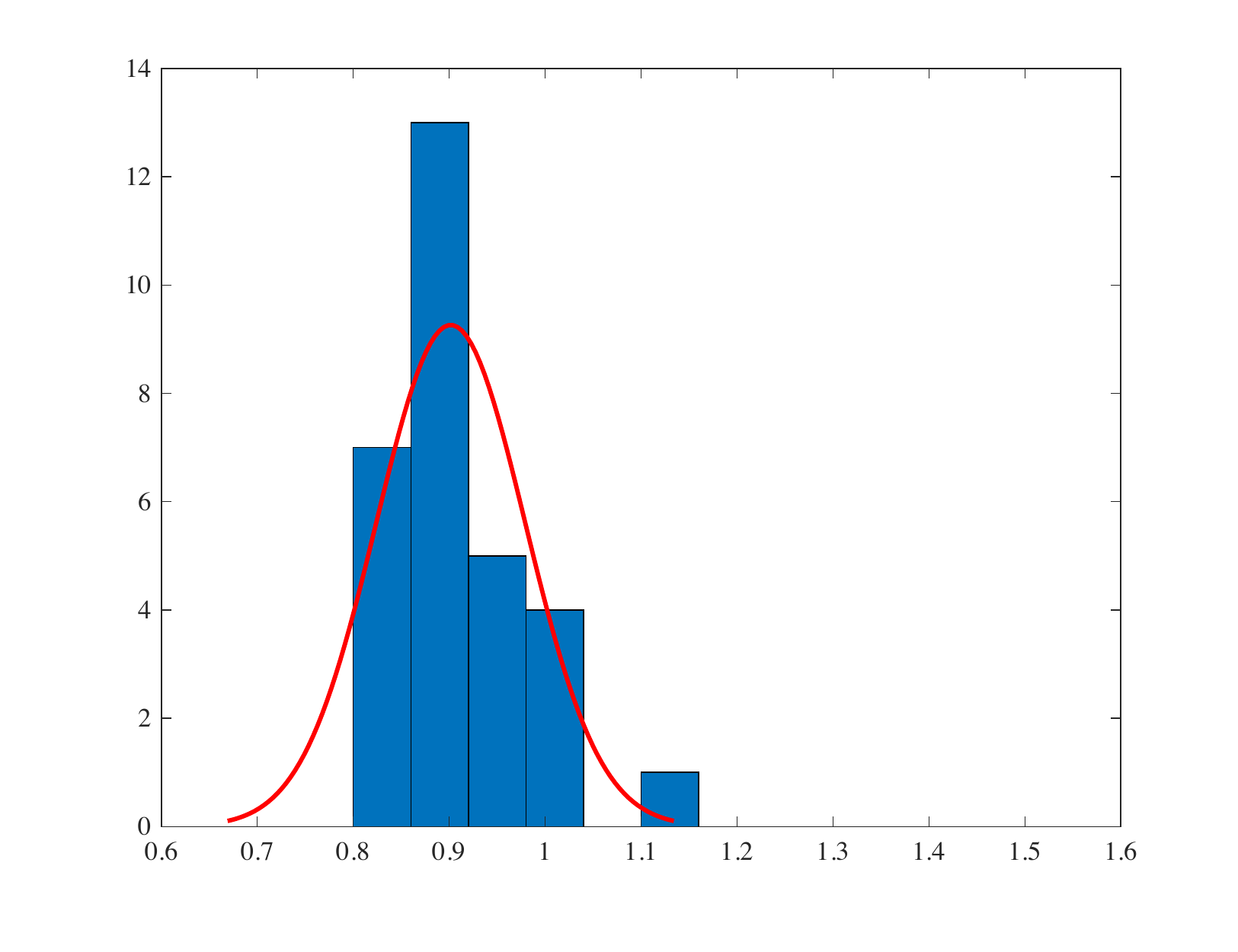}
    \vspace{-3ex}
    \caption{Histogram showing distribution of $E_{min}$ for $30$ runs of Case $II$ using proposed method}
    \label{hist_case2_GA}
    \vspace{-3ex}
\end{figure}

\begin{figure}[h!]
\vspace{-2ex}
    \hspace{-2.2ex}
    \includegraphics[scale=0.5]{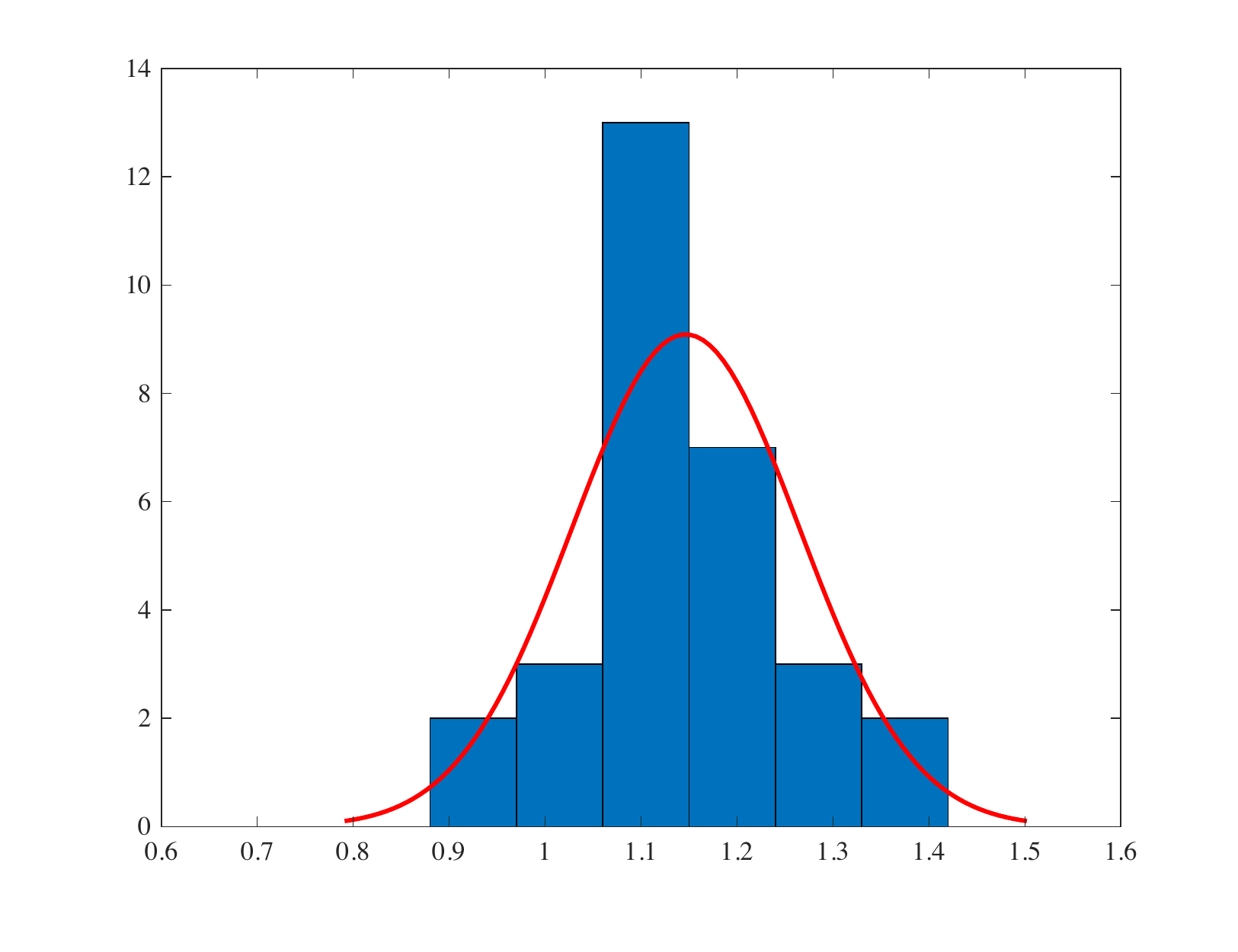}
    \vspace{-2.5ex}
    \caption{Histogram showing distribution of $E_{min}$ for $30$ runs of Case $II$ using stochastic hill climber}
    \label{hist_case2_HC}
    \vspace{-1ex}
\end{figure}

\begin{figure}[h!]
    \centering
    \hspace{-2.2ex}
    \includegraphics[scale=0.5]{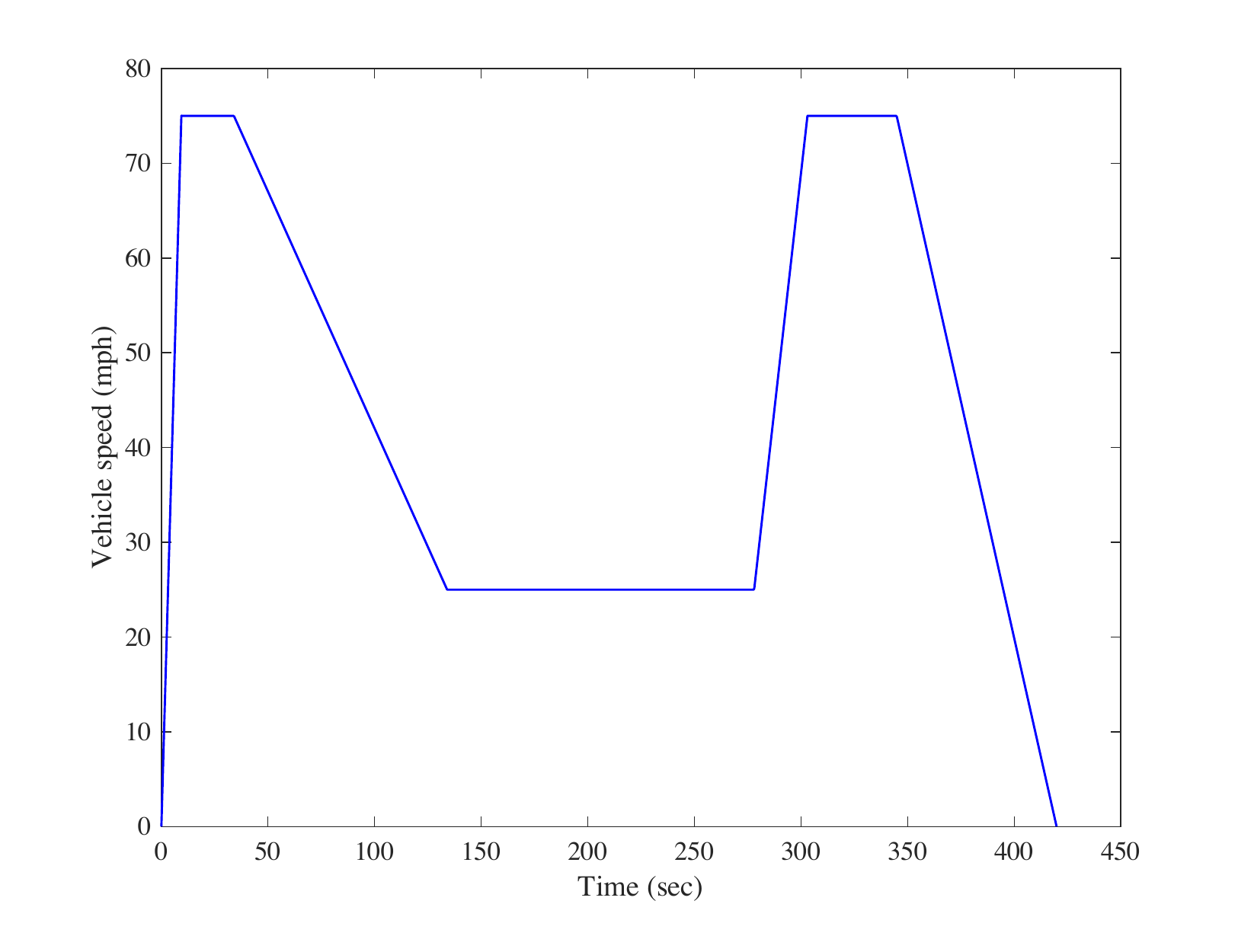}
    \vspace{-2ex}
    \caption{Optimal Speed Profile for Case $II$}
    \label{speedprofile_case2}
    \vspace{-1ex}
\end{figure}

The optimized value of parameters for the proposed method and SHC is shown in Table~\ref{tab:Case2}. Table~\ref{tab:Case2} also shows the overall minimum energy consumption ($E_{min}$) and the average ($E_{avg}$) and standard deviation ($\sigma$) of minimum energy consumption over $30$ runs for both methods. The results in the table show that SHC got stuck over local optimum with $E_{min}$ equal to $0.8851$ kWh whereas the proposed method could search for the better solution with $E_{min}$ equal to $0.8060$ kWh. The table also shows that the results obtained by SHC have higher variance than the proposed method. The histograms showing the distribution of minimum energy consumption ($E_{min}$) over $30$ runs for the proposed method and SHC are shown in Fig.~\ref{hist_case2_GA} and Fig. \ref{hist_case2_HC} respectively. The optimal speed profile obtained for Case $II$ using the proposed method is shown in Fig. \ref{speedprofile_case2}.
\begin{table*}[]
\caption{Results for Case $II$ over 30 runs \vspace{-1.5ex}}
    \centering
        \label{tab:Case2}
\begin{tabular}{|c|c|c|c|c|c|c|c|c|c|c|}
\hline
\multirow{2}{*}{Methods}  & \multicolumn{7}{c|}{Optimized   Parameters} & \multirow{2}{*}{$E_{min}$ (kWh)} & \multirow{2}{*}{$E_{avg}$ (kWh)} & \multirow{2}{*}{$\sigma$ (kWh)} \\ \cline{2-8}
                          & \begin{tabular}[c]{@{}c@{}}$\alpha_1$ \\ (mph/s)\end{tabular} & \begin{tabular}[c]{@{}c@{}}$V_1$ \\ (mph)\end{tabular} & \begin{tabular}[c]{@{}c@{}}$\beta_1$ \\ (mph/s)\end{tabular} & \begin{tabular}[c]{@{}c@{}}$V_2$ \\ (mph)\end{tabular} & \begin{tabular}[c]{@{}c@{}}$\alpha_2$ \\ (mph/s)\end{tabular} & \begin{tabular}[c]{@{}c@{}}$V_3$ \\ (mph)\end{tabular} & \begin{tabular}[c]{@{}c@{}}$\beta_2$ \\ (mph/s)\end{tabular} &                             &                             &                             \\ \hline
                          & & & & & & & &  &&\\
Stochastic   Hill Climber & 4.5                                                   & 75                                                  & 0.5                                                   & 25                                                  & 0.5                                                   & 75                                                  & 4.5                                                   & 0.8851                      & 1.1465                      & 0.1185                      \\
& & & & & & & &  &&\\
Proposed   Method         & 8                                                     & 75                                                  & 0.5                                                   & 25                                                  & 2                                                     & 75                                                  & 1                                                     & 0.8060                      & 0.9016                      & 0.0775                      \\ \hline
\end{tabular}\vspace{-1ex}
\end{table*}

\section{Conclusion}\label{conclusion}
This paper has proposed a GA-based eco-driving technique of EVs considering regenerative braking. The GA was used to determine the optimal or near-optimal combination of the driving parameters that result in minimum energy consumption. The individual chromosomes of the population are represented in binary coding. The proposed method was implemented in two different cases. In the first case, a $5$-mile road without any speed limits is considered, while in the second case a $5$-mile road with a speed limit is considered. The comparison of the results obtained by the proposed method with that of stochastic hill climber shows that the proposed method can effectively determine the optimal or near-optimal solution.   

\bibliographystyle{IEEEtran}
\bibliography{References.bib}

\begin{thebibliography}{10}
\providecommand{\url}[1]{#1}
\csname url@samestyle\endcsname
\providecommand{\newblock}{\relax}
\providecommand{\bibinfo}[2]{#2}
\providecommand{\BIBentrySTDinterwordspacing}{\spaceskip=0pt\relax}
\providecommand{\BIBentryALTinterwordstretchfactor}{4}
\providecommand{\BIBentryALTinterwordspacing}{\spaceskip=\fontdimen2\font plus
\BIBentryALTinterwordstretchfactor\fontdimen3\font minus
  \fontdimen4\font\relax}
\providecommand{\BIBforeignlanguage}[2]{{%
\expandafter\ifx\csname l@#1\endcsname\relax
\typeout{** WARNING: IEEEtran.bst: No hyphenation pattern has been}%
\typeout{** loaded for the language `#1'. Using the pattern for}%
\typeout{** the default language instead.}%
\else
\language=\csname l@#1\endcsname
\fi
#2}}
\providecommand{\BIBdecl}{\relax}
\BIBdecl

\bibitem{NB2018thesis}
N.~Bhusal, ``The combined effect of photovoltaic and electric vehicle
  penetration on conservation voltage reduction in distribution system,'' {MS}
  thesis, University of Nevada, Las Vegas, 2018.

\bibitem{INSIDEEVs}
\BIBentryALTinterwordspacing
M.~Kane. (2020, FEB) Global ev sales for 2019 now in: Tesla model 3 totally
  dominated. [Online]. Available:
  \url{https://insideevs.com/news/396177/global-ev-sales-december-2019/\#:~:text=Around\%202.2\%20million\%20passenger\%20plug,
  which\%20makes\%20us\%20cautiously\%20optimistic.}
\BIBentrySTDinterwordspacing

\bibitem{heydari2020optimal}
S.~Heydari, P.~Fajri, R.~Sabzehgar, and A.~Asrari, ``Optimal brake allocation
  in electric vehicles for maximizing energy harvesting during braking,''
  \emph{IEEE Transactions on Energy Conversion}, 2020.

\bibitem{8378052}
G.~P. {Padilla}, S.~{Weiland}, and M.~C.~F. {Donkers}, ``A global optimal
  solution to the eco-driving problem,'' \emph{IEEE Control Systems Letters},
  vol.~2, no.~4, pp. 599--604, 2018.

\bibitem{7265166}
A.~{Sciarretta}, G.~{De Nunzio}, and L.~L. {Ojeda}, ``Optimal ecodriving
  control: Energy-efficient driving of road vehicles as an optimal control
  problem,'' \emph{IEEE Control Systems Magazine}, vol.~35, no.~5, pp. 71--90,
  2015.

\bibitem{8286942}
J.~{Han}, A.~{Sciarretta}, L.~L. {Ojeda}, G.~{De Nunzio}, and L.~{Thibault},
  ``Safe- and eco-driving control for connected and automated electric vehicles
  using analytical state-constrained optimal solution,'' \emph{IEEE
  Transactions on Intelligent Vehicles}, vol.~3, no.~2, pp. 163--172, 2018.

\bibitem{dongreoptimization}
N.~R. Dongre and A.~Sindekar, ``Optimization of energy consumption in electric
  traction system by using interior point method.''

\bibitem{PAREDES2019556}
J.~F. Paredes, G.~P. Cazar, and M.~Donkers, ``A shrinking horizon approach to
  eco-driving for electric city buses: Implementation and experimental
  results,'' \emph{IFAC-PapersOnLine}, vol.~52, no.~5, pp. 556 -- 561, 2019,
  9th IFAC Symposium on Advances in Automotive Control AAC 2019.

\bibitem{8703854}
D.~{Maamria}, K.~{Gillet}, G.~{Colin}, Y.~{Chamaillard}, and C.~{Nouillant},
  ``Optimal predictive eco-driving cycles for conventional, electric, and
  hybrid electric cars,'' \emph{IEEE Transactions on Vehicular Technology},
  vol.~68, no.~7, pp. 6320--6330, 2019.

\bibitem{KIM2020114254}
Y.~Kim, M.~Figueroa-Santos, N.~Prakash, S.~Baek, J.~B. Siegel, and D.~M. Rizzo,
  ``Co-optimization of speed trajectory and power management for a
  fuel-cell/battery electric vehicle,'' \emph{Applied Energy}, vol. 260, p.
  114254, 2020.

\bibitem{9045962}
K.~M. {So}, P.~{Gruber}, D.~{Tavernini}, A.~E.~H. {Karci}, A.~{Sorniotti}, and
  T.~{Motaln}, ``On the optimal speed profile for electric vehicles,''
  \emph{IEEE Access}, vol.~8, pp. 78\,504--78\,518, 2020.

\bibitem{7115224}
R.~{Zhang} and E.~{Yao}, ``Eco-driving at signalised intersections for electric
  vehicles,'' \emph{IET Intelligent Transport Systems}, vol.~9, no.~5, 2015.

\bibitem{8481449}
L.~{Thibault}, G.~{De Nunzio}, and A.~{Sciarretta}, ``A unified approach for
  electric vehicles range maximization via eco-routing, eco-driving, and energy
  consumption prediction,'' \emph{IEEE Transactions on Intelligent Vehicles},
  vol.~3, no.~4, pp. 463--475, 2018.

\bibitem{ehsani2018modern}
M.~Ehsani, Y.~Gao, S.~Longo, and K.~Ebrahimi, \emph{Modern electric, hybrid
  electric, and fuel cell vehicles}.\hskip 1em plus 0.5em minus 0.4em\relax CRC
  press, 2018.

\bibitem{9067540}
N.~{Bhusal}, M.~{Abdelmalak}, and M.~{Benidris}, ``Optimum locations of
  utility-scale shared energy storage systems,'' in \emph{2019 8th
  International Conference on Power Systems (ICPS)}, Jaipur, India, 2019, pp.
  1--6.

\bibitem{265940}
E.~S.~H. {Hou}, N.~{Ansari}, and {Hong Ren}, ``A genetic algorithm for
  multiprocessor scheduling,'' \emph{IEEE Transactions on Parallel and
  Distributed Systems}, vol.~5, no.~2, pp. 113--120, 1994.

\bibitem{7764202}
Y.~{Tian}, M.~{Benidris}, S.~{Sulaeman}, S.~{Elsaiah}, and J.~{Mitra},
  ``Optimal feeder reconfiguration and distributed generation placement for
  reliability improvement,'' in \emph{2016 International Conference on
  Probabilistic Methods Applied to Power Systems (PMAPS)}, Beijing, China,
  2016, pp. 1--7.

\bibitem{METAWA201775}
N.~Metawa, M.~K. Hassan, and M.~Elhoseny, ``Genetic algorithm based model for
  optimizing bank lending decisions,'' \emph{Expert Systems with Applications},
  vol.~80, pp. 75 -- 82, 2017.

\bibitem{HIASSAT201793}
A.~Hiassat, A.~Diabat, and I.~Rahwan, ``A genetic algorithm approach for
  location-inventory-routing problem with perishable products,'' \emph{Journal
  of Manufacturing Systems}, vol.~42, pp. 93 -- 103, 2017.

\bibitem{IAS2020mukesh}
M.~{Gautam}, N.~{Bhusal}, and M.~{Benidris}, ``A spanning tree-based genetic
  algorithm for distribution network reconfiguration,'' in \emph{2020 IEEE
  Industry Applications Society Annual Meeting}, Detroit, Michigan USA, October
  2020, pp. 1--6.

\end{thebibliography}
\end{document}